\documentclass[a4paper,12pt,one column]{article}
\begin{document}
\title{PRESENT ACCELERATION OF UNIVERSE, HOLOGRAPHIC ENERGY AND BRANS-DICKE THEORY}
\author{B.Nayak$^1$ and L.P.Singh$^2$ \\Department of Physics
\\Utkal University \\Bhubaneswar-751004 \\India \\
$^1$ bibeka@iopb.res.in \\ $^2$ lambodar\_uu@yahoo.co.in }
\date{ }
\maketitle
\begin{abstract}
The present day accelerated expansion of the universe is naturally addressed within the Brans-Dicke theory just by using holographic dark energy model with inverse of Hubble scale as IR cutoff and power law temporal behaviour of scale factor. It is also concluded that if the universe continues to expand, then one day it might be completely filled with dark energy.
\end{abstract}



parameter, transition redshift .
\newpage
\section{INTRODUCTION}
Brans-Dicke(BD) theory[1] is considered as a natural extension of Einstein's general theory of relativity. In BD theory, the gravitational constant becomes time dependent  varying as inverse of a time dependent scalar field which couples to gravity with a coupling parameter $\omega$. One important property of BD theory is that it gives simple expanding solutions[2,3] for scalar field $\Phi(t)$ and scale factor $a(t)$ which are compatible with solar system observations[4,5,6]. The solar system observations[7] also impose lower bound on $\omega$ ($\left|\omega \right| \geq 10^4$). Many of the cosmological problems[8,9,10,1112,13,14] can be sucessfully explained by using the Brans-Dicke theory and it's extended versions.

The finding of SN Ia observations[15] that the universe is currently undergoing accelerated expansion constitutes the most intriguing discovery in observational cosmology of recent years. As a possible theoretical explanation, it is considered that the vacuum energy with negative pressure termed as the dark energy, is responsible for this acceleration. SN Ia observations also provide the evidence of a decelerated universe in the recent past with transition from deceleration to acceleration occuring at redshift $Z_{q=o} \sim 0.5$[16,17]. The Cosmic Microwave Background(CMB) observations support a spatially flat universe as predicted by the inflationary models[18,19]. The simplest candidate of the dark energy is the cosmological constant. However, the unusal large value of the cosmological constant arising out of the spontaneously broken field theoretic vacuum leads to the search for alternative dynamical dark energy models.

Holographic energy[20,21,22,23] has been considered as a candidate for the dynamical dark energy[24]. Since Newton's gravitational constant is rendered dynamical in Brans-Dicke theory as stated above, it is more natural to study cosmological implications of holographic dark energy in Brans-Dicke theory[25,26]. Further, there have been a number of studies involving interaction between holographic dark energy with matter[27,28,29,30,31,32] taking different options like particle horizon, future horizon and Hubble horizon as IR cutoff. We, in this work, present an integrated study involving interacting holographic dark energy in the Brans-Dicke theory taking inverse of the Hubble scale as IR cutoff. We first obtain the equation of state for dark energy and then find the dark energy density parameters in different evolutionary epochs of the Universe. The difference in our approach, coupled with assumed power law behaviour for scale factor and scalar field have enabled us to obtain a parameterised expression for dark energy density in terms of red shift $z$. This, further, relates our work to various attempts[33] trying to obtain empirical parametrisation of dark energy density in terms of $z$ and provides us an empirical estimate of interaction rate $\Gamma$. We next evaluate the deceleration parameter using the equation of state. From the variation of deceleration parameter with redshift, we find that our analysis sucessfully addresses the problem of present acceleration of the universe and determination of crossover $z$ value.
\section{DARK ENERGY DENSITY}  
For a spatially flat FRW universe filled with dust and dark energy, the gravitational field equations in BD theory take the form  
\begin{eqnarray}
 3M_P^2\Phi  [H^2+H  \frac{\dot{\Phi} }{\Phi}-\frac{\omega}{6} \frac{\dot{\Phi}^2}{\Phi^2}] = \rho_x + \rho_m   
\end{eqnarray}
\begin{eqnarray}
2 \frac{\ddot{a} }{a} + H^2+ \frac{\omega}{2} \frac{\dot{\Phi}^2}{\Phi^2} + 2H \frac{\dot{\Phi} }{\Phi} + \frac{\ddot{\Phi} }{\Phi} = -\frac{p_x}{M_P^2\Phi}
\end{eqnarray} 
where $H= \frac{\dot{a} }{a}$  is  the Hubble  parameter,\\ 
$~~~~~~~~~~\rho_x$ = dark energy density,\\ 
$~~~~~~~~~~\rho_m =$ matter energy density \\ 
$~~~~$and $p_x =$ pressure of the dark energy. 
\\The wave equation for Brans-Dicke scalar field has the form,  
\begin{eqnarray}
M_P^2(\ddot{\Phi} + 3H\dot{\Phi})  = \frac{\rho_x + \rho_m - 3 p_x}{2 \omega+3} & .
\end{eqnarray}
It may be noted that equations (1), (2) and (3) together lead to energy conservation equation
\begin{eqnarray*}
(\dot{\rho_x} + \dot{\rho_m})+3H(\rho_x + \rho_m + p_x) = 0 & .
\end{eqnarray*}
Now we use a dark energy model which rests on following three assumptions ; \\
(i) The dark energy density is derived using holographic principle and is given by[34]
\begin{eqnarray*}
\rho_x=3c^2M_P^2 L^{-2}
\end{eqnarray*}
where c is a dimensionless constant of O(1) and L is infrared(IR) cutoff.\\
(ii) IR cutoff is taken as the inverse of the Hubble scale, i.e. $L=H^{-1}$[35,36] . 

So we can write 
\begin{eqnarray}
\rho_x = 3c^2 M_P^2 H^2 & .
\end{eqnarray}
(iii) Matter and dark energy do not conserve separately but they interact with each other and one may grow at the expense of the other.
\\So the energy conservation equation in the presence of dark energy can be written as 
\begin{eqnarray}
\left . \begin{array}{ll}
\dot{\rho_m} + 3H \rho_m=Q \\       
\dot{\rho_x} + 3H(1+\alpha)\rho_x = -Q  \end{array}  \right \}
\end{eqnarray} 
where $Q = \Gamma \rho_x $ with $\Gamma > 0$ is the reaction rate \\
$~~~~$and $ \alpha =\frac{p_x}{ \rho_x}$ denotes the equation of state parameter for the dark energy.
\\Let us assume that the Brans-Dicke field varies with time as a power law of the scale factor $a$ like
\begin{eqnarray}
  \Phi(t) \propto a^n & .
\end{eqnarray}   
Putting equation(6) in equation(1), we get 
\begin{eqnarray}
\rho_x+\rho_m= 3M_P^2 \Phi H^2 [(n+1) - \frac{n^2\omega}{6} ] & .
\end{eqnarray}
Again using equation(4) and the first equation of (5), we can write
\begin{eqnarray}
\dot{\rho_x} = - \rho_x[\frac{(\Gamma-3Hr)-\frac{nH}{c^2}\Phi\{(n+1)-\frac{n^2\omega}{6}\} }{\frac{\Phi}{c^2}\{(n+1)-\frac{n^2\omega}{6}-\frac{c^2}{\Phi}\} }]
\end{eqnarray}
where $r=\frac{\rho_m}{\rho_x}$ is the ratio of matter and dark energy densities.\\
Equations (4) and (7) together lead to,
\begin{eqnarray}
c^2r=\Phi[(n+1)-\frac{n^2\omega}{6}]-c^2
\end{eqnarray}
This equation enables us to obtain a generic value for $n$ as $\sim -0.016$ with both $c$ and $\Phi$ of $O(1)$ and $\left|\omega\right| \cong 10^4$ .\\
Equations (8) and (9),in turn, lead to 
\begin{eqnarray}
\dot{\rho_x}=\rho_x[\frac{\Gamma-3Hr-nH(1+r)}{r}]
\end{eqnarray}
\\By comparing equation(10) with the second equation of (5), an expression for the equation of state parameter of the dark energy is obtained as,
\begin{eqnarray}
\alpha=\frac{n}{3}(1+\frac{1}{r})-\frac{\Gamma}{3H}(1+\frac{1}{r}) & .
\end{eqnarray}
The equation(11) can be written in the form 
\begin{eqnarray}
\alpha=A-\frac{B}{H}
\end{eqnarray}
where 
\begin{eqnarray}
A=\frac{n}{3}(1+\frac{1}{r})\\
B=\frac{\Gamma}{3}(1+\frac{1}{r})  & .
\end{eqnarray}
Further, since $H=-\frac{\dot{z}}{1+z}$ where $z$ is the redshift, one can write\begin{eqnarray}
\alpha(z)=A+(\frac{1+z}{\dot{z}} )B & .
\end{eqnarray}
Using temporal behaviour of the  scale factor $a(t) \propto t^\beta$ with +ve $\beta$ for expanding universe, we get $H=\frac{\beta}{t}$ . Thus,
\begin{eqnarray}
\alpha=A-\frac{B}{\beta} t & .
\end{eqnarray}
Equations (12), (15) and (16) represent different forms of equation of state parameter of the dark energy.

The dark energy density parameter is defined as [33]
\begin{eqnarray}
\left. \begin{array}{lllll}
\Omega_x=\frac{\rho_x}{\rho_c}  & $with$ & \rho_x=\rho_x^0 f(z)  & $and$ & f(z)=exp[3 \int_0^z\frac{1+\alpha(z')}{1+z'} \,dz'] \end{array} \right. & .
\end{eqnarray}
The $z$ dependences of A and B going like $\frac{1}{r} \sim (1+z)^n$ with $n\sim -0.016$ are very weak. So for the purpose of integration in equation(17) to find $f(z)$, A and B can be taken as constants. But for all other purposes, A and B will carry the weak $z$ dependence.\\
Under above assumption equations (15) and (17) lead to 
\begin{eqnarray}
\Omega_x=[1+ \frac{\Omega_m^0}{\Omega_x^0}(1+z)^{-3A} exp\{-3B(t-t_0) \} ]^{-1} & .
\end{eqnarray}
Further, use of values of A and B (equations (13) and (14) ) leads to 
\begin{eqnarray}
\Omega_x=[1+\frac{\Omega_m^0}{\Omega_x^0}(1+z)^{\left| n \right|(1+\frac{1}{r}) }  exp \{ \Gamma(t_0-t)(1+\frac{1}{r}) \} ]^{-1} & .
\end{eqnarray}
\section{ESTIMATION OF $\Gamma$}
Equation (19) can be used for estimation of $\Gamma$ . Since equation (19) leads to $\Omega_x \approx 0$ for $t \approx 0$ and $z \gg 1$, one is inclined to assume a model that the universe started with only relativistic matter and dark energy  appeared due to it's decay with $\Gamma$ as the interaction rate between matter and dark energy. By considering experimental observation that at present time $70 \% $ of the universe is filled with dark energy and rest are matter which has been achieved in $14 \times 10^9 years \approx 1.01 \times H_0^{-1}$ through their interaction, one estimates that
\begin{eqnarray}
\Gamma \approx 5 \times 10^{-11}( yr)^{-1} \approx 0.7 \times H_0 & .
\end{eqnarray} 
\section{$\Omega_x$ FOR DIFFERENT ERA}
\subsection{For radiation dominated era} 
For this era, we take $a(t)\sim t^{\frac{1}{2}}$, since it is well known that any deviation from this behaviour will disturb the primordially formed nuclei abundance in the universe.
Defination of redshift ($\frac{a(t_0)}{a(t)}=1+z$) gives
\begin{eqnarray*}
t= \frac{t_0}{(1+z)^2}  & .
\end{eqnarray*}
Putting this in equation(19), one gets 
\begin{eqnarray}
\Omega_x=[1+\frac{\Omega_m^0}{\Omega_x^0}(1+z)^{\left| n \right|(1+\frac{1}{r})} exp \{\Gamma t_0(1-\frac{1}{(1+z)^2} )(1+\frac{1}{r}) \}]^{-1} & .
\end{eqnarray}
\subsection{For matter dominated era  }
For this era, we have $a(t)\propto t^{(2\omega+2)/(3\omega+4)}\approx t^\frac{2}{3} $ .\\  
Defination of redshift gives
\begin{eqnarray*}
t=\frac{t_0}{(1+z)^\frac{3}{2} } & .
\end {eqnarray*}
Inserting this in equation(19), one gets
\begin{eqnarray}
\Omega_x=[1+\frac{\Omega_m^0}{\Omega_x^0}(1+z)^{\left| n \right|(1+\frac{1}{r})} exp \{ \Gamma t_0(1-\frac{1}{(1+z)^\frac{3}{2}})(1+\frac{1}{r})  \} ]^{-1} & .
\end{eqnarray}

Using equations (12), (20), (21) and (22), we can get \\
(i) in the distant past for $z \gg 1$ and $t \approx 0$, $ \alpha \to -0.005$ and $\Omega_x \to 0$\\
(ii) in the distant future for $(1+z) \to 0$ and $t \gg 1$, $\alpha \to -\infty$ and $\Omega_x \to 1$ .

Further, these equations satisfy following  observational constraints[37,38,39]
\\(a)Last Scattering Surface(LSS) Constraint :  During the galaxy formation era ($1<z<3$)  dark energy density must be sub-dominant to matter density, accordingly $\Omega_x< 0.5$ . Equation(22) yields $0.24\leq \Omega_x \leq 0.34$ for galaxy formation era. 
\\(b)Big Bang Nucleosynthesis(BBN) Constraint : The presence of dark energy in nucleosynthesis era, should not disturb the observed Helium abundance in the universe which is regarded as one of the biggest support of Big Bang Theory. Cybrut found that $(\Omega_x)_{BBN}<0.21$ at $ z=10^{10}$ . Our equation(21) yields $(\Omega_x)_{BBN} \sim 0.227$ which is quite close to the experimental bound. 
\subsection{Present value of $\alpha$ and $\Omega_x$ }
Kaplinghat et al [40] and others [41] have pointed out that for power law cosmologies, high redshift data and present age of the universe restricts $\beta$ to value $\approx 1$ . Taking $\beta=1$ and using equation (16), we get $\alpha_0=0.8$ . But accelerated expansion of the universe requires $\beta$ to be greater than $1$ though exact value of $\beta$ has not been ascertained. Using $\beta=1+\epsilon$ and comparing with the experimental value of $\alpha_0$ (i.e. $\alpha_0 \leq -0.72$ ), we get $\epsilon \leq 0.05$ . For a typical $\beta=1.01$, one finds $\alpha_0=-0.79$ . 
\\Putting $z=0$ for present era in equation(22), we get 
\begin{eqnarray*}
\Omega_x^0=\frac{\Omega_x^0}{\Omega_x^0+\Omega_m^0}
\end{eqnarray*}
which implies
\begin{eqnarray*}
\Omega_x^0+\Omega_m^0=1
\end{eqnarray*}
in confirmity with inflationary paradigm and experimental finding of the universe being flat.
\section{DECELERATION PARAMETER $(q)$ FOR DIFFERENT ERA}
Dividing equation(2) by $H^2$ and using equations (4) and (6), we can get an expression for deceleration parameter as	 
\begin{eqnarray*}
q=\frac{1}{n+2}[\frac{3\alpha (z)c^2}{\Phi}+\frac{n^2 \omega}{2}+(n^2+n+1)] 
\end{eqnarray*}
So for distant past($z \to \infty$) $q \to 1.17$ and for distant future($1+z \to 0$) $q \to -\infty$.
Further, for present era($z=0$) $q \approx -0.025$ indicating accelerated expansion. The transition redshift from decelerated expansion to present acceleration is obtained as $z_{q=0} \sim 0.32$ which is in good agreement with experimental observation of $z_{q=0} \sim 0.5$. 
\section{COINCIDENCE PROBLEM}
From energy conservation equation (5), we get 
\begin{eqnarray}
\frac{\dot{r}}{r}=3H[\alpha+(1+\frac{1}{r})\frac{\Gamma}{3H}]
\end{eqnarray}
Using current values for various parameters in the above equation, we find 
\begin{eqnarray}
\left|\frac{\dot{r}}{r}\right|_0=2.0 \times 10^{-2} \times 3H_0
\end{eqnarray}
So $r$ varies more slowly in this model than in the conventional $\Lambda$CDM model where $\left|\frac{\dot{r}}{r}\right|_0=3H_0$ . Thus, the coincidence problem gets more softened in the present case.

\section{DISCUSSION AND CONCLUSION}
We first note that our integrated model involving interacting holographic dark energy based on Brans-Dicke theory can accomodate the present value of dark energy equation of state parameter $\alpha$ which is within the experimental bounds. Further, the negative value of $q$ for present era indicates the accelerating expansion of the universe. We obtain the transition redshift from decelerated to accelerated expansion is $z_{q=0} \sim 0.32$ in fairly good agreement with SN Ia observation. We find that the expansion of the universe remains accelerating for ever. 

The calculated values of dark energy density parameter for different era in the history of expansion is fairly consistent with LSS and BBN constraints. In the context of our model this implies that in the distant past universe was filled with matter and in future universe will completely be filled with dark energy if it continues to expand. This, incidentally, is the scenario discussed by M.Li, C.Lin and Y.Wang[42].

Brans-Dicke theory in conjuction with the notion of holographic dark energy, thus, provides a satisfactory description of various experimental observational facts like satisfying constraints on dark energy density, flat universe and it's present acceleration including the crossover value of $z$ . It  also considerably softens the coincidence problem.
\section*{ACKNOWLEDGMENTS}
We are thankful to Institute of Physics, Bhubaneswar, India, for providing the library and computational facility. B.Nayak would like to thank the Council of Scientific and Industrial Research, Government of India, for the award of JRF, F.No. 09/173(0125)/2007-EMR-I .  
\section*{REFERENCES}
$[1]$ C. H. Brans and R. H.  Dicke, Phys. Rev. $\textbf{124}$, $925$ ($1961$) .\\
$[2]$ C. Mathiazhagan and V. B. Johri, Class. Quantum Grav. $\textbf{1}$, $L29$ ($1984$) .\\
$[3]$ D. La and P. J. Steinhardt, Phys. Rev. Lett $\textbf{62}$, $376$ ($1989$) .\\
$[4]$ S. Perlmutter et al., Astrophys.J. $\textbf{517}$, $565$ ($1999$) .\\
$[5]$ A. G. Riess et al., Aston.J. $\textbf{116}$, $74$ ($1999$) .\\ 
$[6]$ P. M. Garnavich et al., Astrophys. J. $\textbf{509}$, $74$ ($1998$) .\\
$[7]$ B. Bertotti, L. Iess and P. Tortora, Nature $\textbf{425}$, $374$ ($2004$) .\\
$[8]$ D. La, P. J. Steinhardt and E. W. Bertschinger, Phys. Lett. B $\textbf{231}$, $231$ ($1989$) .\\
$[9]$ M. C. Bento, O. Bertolami, and P. M. Sa, Phys. Lett. B $\textbf{262}$, $11$ ($1991$) .\\
$[10]$  S. J. Kolitch and D. M. Eardly, Ann. Phys. (N.Y.) $\textbf{241}$, $128$ ($1995$) .\\
$[11]$ J. D. Barrow et al., Phys. Rev. D $\textbf{48}$, $3630$ ($1993$) .\\
$[12]$ J. D. Barrow et al., Mod. Phys. Lett. A $\textbf{7}$, $911$ ($1992$) .\\
$[13]$ B. K. Sahoo and L. P. Singh, Mod. Phys. Lett. A $\textbf{17}$,  $2409$ ($2002$) .\\
$[14]$ B. K. Sahoo and L. P. Singh, Mod. Phys. Lett. A $\textbf{18}$, $2725$ ($2003$) .\\
$[15]$ A. G. Riess et al., Astrophys. J. $\textbf{607}$, $665$ ($2004$) .\\
$[16]$ M. S. Turner and A. G. Riess, Astrophys. J. $\textbf{569}$, $18$ ($2002$) .\\
$[17]$ A. G. Riess, Astrophys. J. $\textbf{560}$, $49$ ($2001$) .\\
$[18]$ de P. Bernardis et al., Nature $\textbf{404}$, $955$ ($2000$) .\\ 
$[19]$ S. Hanany et al., Astrophys. J. $\textbf{545}$, $L5$ ($2000$) .\\
$[20]$ G.'t Hooft,gr-qc$/9310026$ .\\
$[21]$ L. Susskind, J. Math. Phys.(N.Y.) $\textbf{34}$, $6377$ ($1995$) .\\ 
$[22]$ A. G. Cohen, D. B. Kaplan and A. E. Nelson, Phys. Rev. Lett. $\textbf{82}$, $4971$ ($1999$)\\
$[23]$ M. Li, Phys. Lett. B $\textbf{603}$, $1$ ($2004$) .\\
$[24]$ R. Horvat, Phys. Rev. D $\textbf{70}$ , $087301$ ($2004$) .\\
$[25]$ Hungoo Kim, H. W. Lee and Y. S. Myung, Phys. Lett. B $\textbf{628}$, $11$ ($2005$) .\\
$[26]$ Yungui Gong, Phys. Rev. D $\textbf{70}$, $064029$ ($2004$) .\\
$[27]$ Bin Wang, Yungui Gang and Elcio Abdalla, Phys. Lett. B $\textbf{624}$, $141$ ($2005$) .\\ 
$[28]$ Hungsoo Kim, H. W. Lee and Y. S. Myung, Phys. Rev. d $\textbf{632}$,($2006$) $605$ .\\
$[29]$ Q. Wu, Y. Gong, A. Wang and J. S. Alcaniz, Phys. Lett. B $\textbf{659}$, $34$ ($2008$) .\\
$[30]$ R. Horvat, Phys. Rev. D $\textbf{70}$, $087301$ ($2004$) .\\
$[31]$ Narayan Banerjee and Diego Pavon, Phys. Lett. B $\textbf{647}$, $477$ ($2007$) .\\
$[32]$ D. Pavon, J. Phys. A $\textbf{40}$, $6865$ ($2007$) .\\
$[33]$ V. B. Johri and P. K. Rath, Int. J. Mod. Phys. D $\textbf{16}$, $1581$ ($2007$) .\\
$[34]$ M. Li, Phys. Lett. B $\textbf{603}$, $1$ $(2004)$ .\\
$[35]$ Masata Ito, Europhys. Lett. $\textbf{71}$, $712$ ($2005$) .\\
$[36]$ A. G. Cohen, D. B. Kaplan and A. E. Nelson, Phys. Rev. Lett. $\textbf{82}$, $4971$ $(1999)$\\
$[37]$ W. L. Freedman and M. S. Turner, Rev. Mod Phys. $\textbf{75}$, $1433$ ($2003$) .\\
$[38]$ V. B. Johri, Pramana $\textbf{59(3)}$, $L553-561$ ($2002$)  .\\
$[39]$ R. H. Cyburt, B. D. Fields, K. A. Olive and E. Skillman, Astropart. Phys.$\textbf{23}$, 

$313$ ($2005$) .\\
$[40]$ M. Kaplinghat, G. Steigman, I. Tkachev and T. P. Walker, Phys. Rev. D $\textbf{59}$, 

$043510$ ($1999$) .\\
$[41]$ M. Sethi, A. Batra and D. Lohiya, Phys. Rev. D $\textbf{60}$, $108301$ ($1999$) .\\
$[42]$ M. Li, C. Lin and Y. Wang, JCAP $\textbf{05}$, $023$ ($2008$) .
\end{document}